\documentclass[a4paper, 12pt]{article}
\usepackage{amsfonts}
\usepackage{epsfig}
\usepackage{graphicx}

\begin{document}

\begin{center}
\fontsize{14pt}{16pt} \bf On the solutions of the critical Lane--Emden equation in higher space dimensions
\end{center}

\begin{center}
\sf{Rados\l aw Antoni Kycia$^{1}$, Galina Filipuk$^{2}$ }
\end{center}

\medskip
\small{
\centerline{ University of Warsaw}
\centerline{ Faculty of Mathematics, Informatics and Mechanics}
\centerline{ Banacha 2, Warsaw, 02-097, Poland}
 \centerline{$^{1}${\tt
kycia.radoslaw@gmail.com}, $^{2}${\tt
filipuk@mimuw.edu.pl}}}
\bigskip

\begin{quote} {\bf Abstract.}
{\it  In this paper we study   solutions of the critical Lane--Emden  equation in higher space dimensions. We show that after certain transformations the general solution can be written in terms of elliptic functions. We restrict ourselves to real solutions which can be used in physical applications.}
\end{quote}

\bigskip
\normalsize

\section{Introduction}

The computer algebra systems, for instance {\it Mathematica} \cite{math}, are useful for making numeric and symbolic computations for ordinary differential equations and to visualize the results.

The Lane--Emden (LE) equation \cite{Mathematica_Lane-Emden} is one of the most important classical equations of mathematical physics. It originally appeared in astrophysics. It originates from studying static star structures and was proposed by Lane in \cite{Lane} and Emden in \cite{Emden}. Later on, it also appeared in kinetic theory, quantum mechanics and other fields (see \cite{Mach_Lane_Emden, Havas} and the references therein).

The LE equation is of the form
\begin{equation}
 \triangle \theta + \theta^{p}=0,
\label{Lane-Emden_general}
\end{equation}
where $p$ is a   natural number. Here $\theta=\theta(x_1,\ldots,x_d)$ and $d$ is space dimension.  To preserve the reflection symmetry the nonlinear term in (\ref{Lane-Emden_general}) is usually rewritten as $|\theta|^{p-1}\theta$, however we will not consider this case. Equation (\ref{Lane-Emden_general}) appears in many physical and mathematical applications even in higher than three space dimensions, see, for instance,  \cite{Chandrasekhar}, \cite{Benguria1}, \cite{Benguria2}, \cite{Kycia_PDE} and the references therein.

If the spherical symmetry is assumed, equation (\ref{Lane-Emden_general}) can be formulated on the positive semiline as
\begin{equation}
 \theta '' +\frac{d-1}{x}\theta'+\theta^{p}=0,
 \label{Lane-Emden_spherical}
\end{equation}
where $\theta=\theta(x)=\theta\left(\sqrt{\sum_{i=1}^d x_i^2}\right)$ and $'=d/dx$. This equation possesses the scaling symmetry, i.e., the  scaled solution
\begin{equation}
 \theta_{\lambda}(x)=\frac{1}{\lambda^{\alpha}}\theta\left(\frac{x}{\lambda}\right),\qquad \alpha=\frac{2}{p-1}
\label{scaling}
\end{equation}
is   a solution as well. It enables us to generate solutions with different initial conditions from  the normalized ones.

The simplest solution is of the form
\begin{equation}
 \theta(x)=b_{\infty}\,x^{-\alpha}, \qquad b_{\infty}= \left(\frac{2(p(d-2)-d)}{(p-1)^{2}}\right)^{1/(p-1)}.
\label{b_solution}
\end{equation}
It can be obtained by assuming a power type form for $\theta(x)$, substituting it into the equation (\ref{Lane-Emden_spherical}) and equating the coefficients.  The solution is singular at the origin and, therefore, it is of no direct physical importance, however it is of crucial importance in our further analysis. Other closed form solutions are known only for a few special values of $d$ and $p$.  In general, solutions with the typical initial data $\theta(0)=1$, $\theta'(0)=0$ can be obtained in terms of the power series and they are well known. The discussion of movable singularities of solutions can be found in  \cite{Lane_Emden_Hunter} and \cite{Lane_Emden_szeregi}. We also note that this equation falls in the class of equations considered in \cite{GF_RH}.

One of the nontrivial cases when the solution can be found in a closed form is the critical case when
\begin{equation}
 p_{Q}=:p=\frac{d+2}{d-2}.
\label{critical_p}
\end{equation}
Note that it is the case when a functional, called the energy functional in physical applications,
\begin{equation}
 E[\theta]=\int d^{d}x \left(\frac{1}{2}\theta'^{2}+\frac{1}{p+1}\theta^{p+1}\right)
\end{equation}
is scale invariant under (\ref{scaling}). The closed form solution for this case generalizes the Schuster and Emden solution for $d=3$ and it is sometimes   called the generalized Talenti-Aubin solution, especially when one considers nonlinear wave equations \cite{Kycia_PDE}. It is of the form
\begin{equation}
 \theta(x)=\frac{1}{(1+ax^{2})^{\alpha}},\qquad a=\frac{p-1}{4d}
 \label{Talenti-Aubin_solution}
\end{equation}
where $p$ and $d$ satisfy  (\ref{critical_p}).

\section{Main result}

The Lane--Emden equation can be analyzed by the Emden substitution \cite{Emden}
\begin{equation}
 \theta=z\,x^{-\alpha},\qquad y=-\ln(x),
\label{Emden_substitution}
\end{equation}
which gives the equation
\begin{equation}
 z''+\frac{2p-dp+d+2}{p-1}z'+\frac{2(2-d)p+2d}{(p-1)^{2}}z+z^{p}=0,
\label{equation_Emden_substitution}
\end{equation}
where $z=z(y)$ and  $'=d/dy$. However, a slight modification of this substitution which generalizes \cite{Mach_Lane_Emden} in the form
\begin{equation}
 \theta=b_{\infty}\,z\,x^{-\alpha},\qquad y=-\ln(x),
\label{Modified_Emden_substitution}
\end{equation}
where $b_{\infty}$ is given by (\ref{b_solution}), symmetrizes the last two terms and we obtain
\begin{equation}
 z''+\frac{2p-dp+d+2}{p-1}z'+\frac{2(2-d)p+2d}{(p-1)^{2}}(z-z^{p})=0.
\label{equation_Modified_Emden_substitution}
\end{equation}
Both these equations can be interpreted as equations describing a particle moving with friction (the term with $z'$) in the field of forces given by the last two terms. For the later analysis we consider    equation (\ref{equation_Modified_Emden_substitution}).
It is surprising that for the critical case the friction term vanishes and we get
\begin{equation}
 z''-\frac{(d-2)^{2}}{4}(z-z^{p})=0,
\label{critical_equation_Modified_Emden_substitution}
\end{equation}
where $p$ and $d$ are connected by (\ref{critical_p}). Due to the lack of the dissipation the energy functional for (\ref{critical_equation_Modified_Emden_substitution}) is preserved \cite{Benguria2}.

The standard method of integration (by multiplying   equation (\ref{critical_equation_Modified_Emden_substitution}) by $2z'$ and integrating) gives
\begin{equation}
 (z')^{2}=\frac{(d-2)^{2}}{2}\left(\frac{1}{2} z^{2}-\frac{1}{p+1} z^{p+1}+C\right),
\label{critical_equation_integrated}
\end{equation}
where $C$ is a constant of integration.  Thus, we see that the general solution of the LE equation in the critical case can be written by using the elliptic functions.

From the physical viewpoint we can only consider  critical cases: (i) $d=3,\;p=5$; (ii) $d=4,\;p=3$ and (iii) $d=6,\;p=2$, and we should analyse only real solutions of equation (\ref{critical_equation_Modified_Emden_substitution}). The real solutions fall into the following classes (see \cite{Benguria2} for a simple proof): (1) positive or negative but not constant, or (2) sign changing, or (3) identically zero, or (4) the solution (\ref{b_solution}) and its negative counterpart when $p$ is odd in case the the nonlinear term in (\ref{Lane-Emden_general}) is rewritten as $|\theta|^{p-1}\theta$.

The case (i) is fully studied in \cite{Mach_Lane_Emden}. Our attention is focused here on cases (ii) and (iii). The motivation to study  such cases is not only for the mathematical completeness, but also for the physical interpretation: the list of all solutions for the critical LE equation enables us to compare all critical cases and answer the question  whether our three dimensional space is distinguished or not.

The form of the real solutions for (\ref{critical_equation_integrated}) depends on the number of zeros of the polynomial on the right hand side  of the equation.  In addition, all roots are constant solutions. Therefore, by (\ref{Modified_Emden_substitution}) they correspond to the solution (\ref{b_solution}).

\section{$d=4$, $p=3$}

In this case the equation (\ref{critical_equation_integrated}) has the form
\begin{equation}
 (z')^{2} = \frac{1}{2}( -z^{4} + 2 z^{2} + C ),
 \label{critical_equation_integrated_d=4_p=3}
\end{equation}
where $C$ is a new constant denoted here for simplicity by $C$.
Now the solution can be obtained by integrating
\begin{equation}
 \pm \int \frac{dy}{\sqrt{2}}=\int \frac{dz}{\sqrt{-z^{4} + 2 z^{2} + C}}.
 \label{differential_equation_d=4_p=3}
\end{equation}

The further analysis relies on the nonnegativeness of the polynomial
\begin{equation}
 w_{43}(z)= -z^{4} + 2 z^{2} + C.
 \label{polynomial_d=4_p=3}
\end{equation}
A few different cases can be distinguished:
\begin{itemize}
 \item { $C < -1$: In this case $w_{43}(z) <0 $ for all real $z$ and, therefore, there are no real solutions.}
 \item { $C = -1$: there are only two points $z=\pm 1$ for which $w_{43}(z)=0$, and they give the solutions that correspond to solution (\ref{b_solution});}
 \item { $ C \in (-1;0)$: there are two disjoint sets on which $w_{42}(z) \ge 0$, namely $$z \in \left[-\sqrt{1+\sqrt{1+C}};-\sqrt{1-\sqrt{1+C}}\right] \cup \left[\sqrt{1-\sqrt{1+C}};\sqrt{1+\sqrt{1+C}}\right]; $$}
 \item { $C = 0$: In this case $w_{43}(z) \ge 0$ when $z \in [-\sqrt{2}; \sqrt{2}]$;}
 \item { $C > 0$: for every $ |z| < z_{0}$, where $z_{0}$ is positive real root  of $w_{43}(z)=0$. In this case there are two real solutions and two imaginary ones.}
\end{itemize}

\begin{figure}[htp]
\centering
 \includegraphics[scale=0.7]{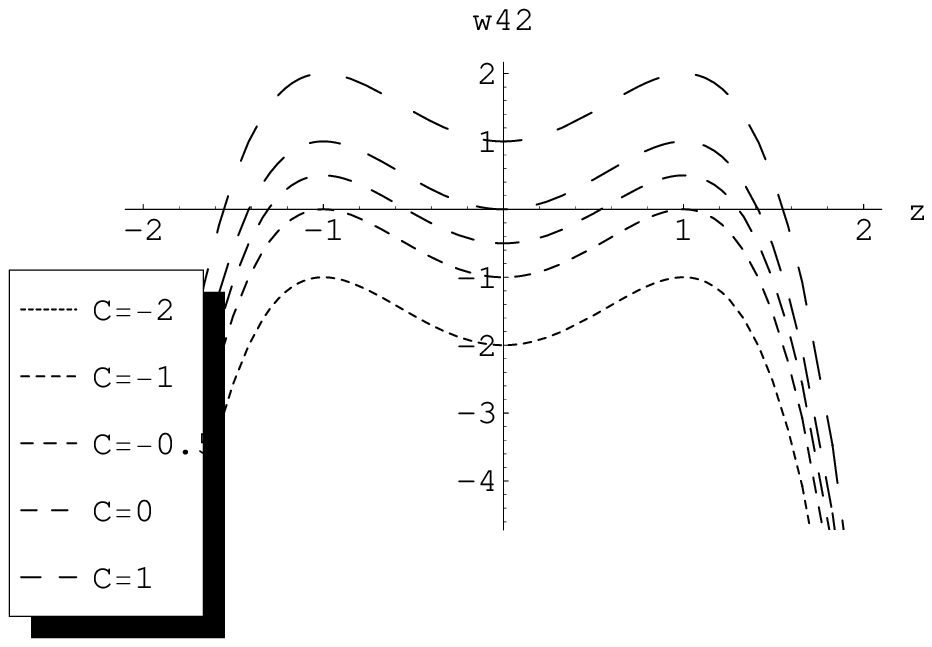}
 \caption{The polynomial (\ref{polynomial_d=4_p=3}) for different values of $C$. Note the set of $z$ where $w_{43}(z) \ge 0$.}
 \label{rysunek_w43_polynomial}
\end{figure}

The Figure \ref{rysunek_w43_polynomial} illustrates the situation.

All the cases will be analysed one by one in the following subsections.

\subsection{$ C \in (-1;0)$}
Let us focus on the case when $z \in \left[\sqrt{1-\sqrt{1+C}};\sqrt{1+\sqrt{1+C}}\right] $. The remaining case can be analysed in a similar way. Equation (\ref{differential_equation_d=4_p=3}) can be factorised as follows:
\begin{equation}
 \pm \int \frac{dy}{\sqrt{2}} = hh' \int \frac{dz}{\sqrt{- \left( 1 - h^{2} z^{2} \right) \left( 1 - h'^{2} z^{2} \right) } },
\end{equation}
where
$$ h =\frac{1}{ \sqrt{1-\sqrt{1+C}} }  >  h'=\frac{1}{ \sqrt{1+\sqrt{1+C}}} > 0.$$
The integral
\begin{equation}
 I=hh'\int_{\frac{1}{h}}^{z_{0} \leq \frac{1}{h'} }   \frac{dz}{\sqrt{- \left( 1 - h^{2} z^{2} \right) \left( 1 - h'^{2} z^{2} \right) } }
\end{equation}
can be brought to the Jacobian elliptic integral \cite{Wang_Guo}, \cite{NIST_math_functions} by the standard substitution \cite{Fichtenholtz}, \cite{Wang_Guo}:
\begin{equation}
h' z =\sqrt{1-\frac{h^{2}-h'^{2}}{h^{2}}u^{2}},
\label{substitution_Fichtengoltz_C_(-10)}
\end{equation}
where now $0 < u <1$. The substitution gives
$$I=h' \int_{0}^{u_{0}} \frac{du}{\sqrt{ ( 1-u^{2}) (1-k^{2}u^{2}) }  }=h' arcsn(u_{0},k),$$
where the elliptic modulus is of the form
$$k=\sqrt{\frac{h^{2}-h'^{2}}{h^{2}}} = \sqrt{\frac{2\sqrt{C+1}}{1+\sqrt{C+1}}},$$
$u_{0}(z_{0})$ can be obtained from (\ref{substitution_Fichtengoltz_C_(-10)}) and $arcsn$ is the inverse of the Jacobian elliptic function $sn()$.
Returning to the original variables $\theta$, $x$ we get the solution
\begin{equation}
\begin{array}{l}
 \theta(x) = \pm\frac{1}{x}\sqrt{\sqrt{1+\sqrt{1+C}} (1-k^{2}y^{2}(x))}, \\
 y(x)=sn(\pm\sqrt{1+\sqrt{1+C}} \frac{\ln(Bx)}{\sqrt{2}},k),
\end{array}
\end{equation}
where $B$ is a constant of integration.

\subsection{$C=0$}
In this case the Talenti-Aubin solution can be recovered as follows. From
$$\pm \int \frac{dy}{\sqrt{2}}=\int \frac{dz}{\sqrt{-z^{4} + 2 z^{2}}}$$
after some simple manipulations one gets
$$z=\sqrt{2}sech(\pm \ln(Bx) )=\pm 2\sqrt{2} Bx \frac{1}{1+(Bx)^{2}},$$
where, as previously, $B$ is an integration constant and $y=-\ln(x)$. Substituting  $2\sqrt{2}B =\frac{1}{\lambda}$ and $\theta (x) = \frac{z}{x}$ we obtain the scaled (see (\ref{scaling})) Talanti-Aubin solution (\ref{Talenti-Aubin_solution})
$$\theta(x)=\pm \frac{1}{\lambda} \frac{1}{1+\frac{1}{8}(x/\lambda)^{2}}.$$

\subsection{$C>0$}
In this case the Jacobian elliptic integrals can also be used. Equation (\ref{differential_equation_d=4_p=3}) can be rewritten in the form
\begin{equation}
 \pm \int \frac{dy}{\sqrt{2}} = hh' \int \frac{dz}{\sqrt{ \left( 1 - h^{2} z^{2} \right) \left( 1 + h'^{2} z^{2} \right) } },
\end{equation}
where now
$$ 0< h =\frac{1}{ \sqrt{1+\sqrt{1+C}} }  <  h'=\frac{1}{ \sqrt{\sqrt{1+C}-1}}.$$
The standard change of variable \cite{Fichtenholtz}
$$hz=\sqrt{1-u^{2}}$$
in the integral
$$I=hh' \int_{\frac{1}{h}}^{z_{0} \leq \frac{1}{h}} \frac{dz}{\sqrt{ \left( 1 - h^{2} z^{2} \right) \left( 1 + h'^{2} z^{2} \right) } }$$
gives ($0 < u \leq 1$)
$$I=-\frac{hh'}{\sqrt{h^{2}+h'^{2}}} \int_{0}^{u_{0}} \frac{du}{\sqrt{(1-u^{2})(1-k^{2}u^{2})}}=-\frac{hh'}{\sqrt{h^{2}+h'^{2}}} arcsn(u_{0},k),$$
where the elliptic modulus is defined by
$$k=\sqrt{\frac{h'^{2}}{h^{2}+h'^{2}}}=\sqrt{\frac{1+\sqrt{C+1}}{2\sqrt{C+1}}}.$$
Therefore, denoting  an integration constant by $B$, we obtain
$$\pm \frac{\ln(Bx)}{\sqrt{2}}=-\frac{1}{\sqrt{2\sqrt{C+1}}}arcsn(u_{0}(z_{0}),k),$$
which gives
\begin{equation}
\begin{array}{l}
 \theta(x)=\pm \frac{1}{x} \sqrt{(1+\sqrt{C+1})(1-y^{2}(x))}, \\
 y(x)=sn(\pm\sqrt{2\sqrt{C+1}}\frac{\ln(Bx)}{\sqrt{2}},k).
\end{array}
\end{equation}

\section{$d=6$, $p=2$}
In this case the equation (\ref{critical_equation_integrated}) is of the form
\begin{equation}
 (z')^{2} = \frac{4}{3}( -2z^{3} + 3 z^{2} + C )
 \label{critical_equation_integrated_d=6_p=2}
\end{equation}
and we obtain
\begin{equation}
  \frac{dz}{\sqrt{-2z^{3} + 3 z^{2} + C}} = \pm \frac{2dy}{\sqrt{3}}.
 \label{differential_equation_d=6_p=2}
\end{equation}

The nonnegativeness of the polynomial
\begin{equation}
 w_{62}(z)= -2z^{3} + 3 z^{2} + C
 \label{polynomial_d=6_p=2}
\end{equation}
determines possible solutions for different $C$ values.

We have to consider the following cases
\begin{itemize}
 \item { $C<-1$: the polynomial $w_{62}(z) \geq 0 $ for $z < z_{0}$, where $z_{0}$ is the only real solution of $w_{62}(z)=0$;}
 \item { $C=-1$:  we have the real root $z_0$ of $w_{62}(z) = 0$ such  that for $z < z_{0}< 0$ we get $w_{62}(z) \geq 0$, and the second real root at $z=1$; }
 \item { $C \in (-1;0)$: there are three real roots $a<b<c$ of $w_{62}(z)=0$, therefore for $z \in (-\infty,a] \cup [b,c]$ the polynomial $w_{62}(z)$ is nonnegative;}
 \item {$C=0$: we have exactly solution (\ref{Talenti-Aubin_solution});}
 \item {$C>0$: there is only one real root $z_{0}$ of $w_{62}(z)=0$, and $w_{62}(z)$ is nonnegative for $z <z_{0}$.}
\end{itemize}

\begin{figure}[htp]
\centering
 \includegraphics[scale=0.7]{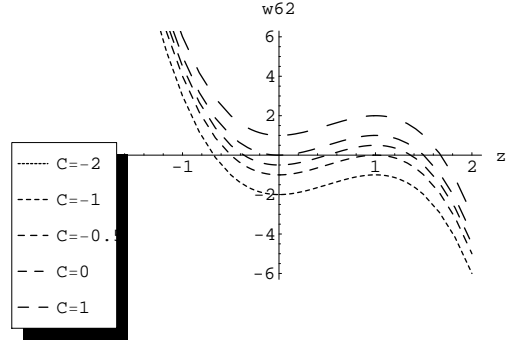}
 \caption{The polynomial (\ref{polynomial_d=6_p=2}) for different values of $C$. Note the set of $z$ where $w_{62}(z) \ge 0$.}
 \label{rysunek_w62_polynomial}
\end{figure}

The cases above can be  illustrated by the plot (\ref{rysunek_w62_polynomial}).

In the following analysis the Weierstrass representation of elliptic integrals \cite{Wang_Guo} will be used. However, this representation can be easily transformed into the Jacobian elliptic integral form using transformations described in \cite{Wang_Guo}.

\subsection{$C<-1$, $C >0$ and $C=1$}
By integrating (\ref{differential_equation_d=6_p=2}) we get the equation
\begin{equation}
 \pm \frac{2}{\sqrt{3}} y + B = \int_{-\infty}^{z<z_{0}} \frac{dz}{\sqrt{-2z^{3} + 3 z^{2} + C}},
 \label{integral_equation_d=6_p=2_case_1}
\end{equation}
where $B$ is an integration constant. The integral on the right-hand side  of the (\ref{integral_equation_d=6_p=2_case_1}) can be brought to the standard form of the Weierstrass elliptic integral \cite{Wang_Guo}, \cite{NIST_math_functions}
$$z=\int_{\infty}^{\xi}\frac{du}{\sqrt{4u^{3}-g_{2}u-g_{3}}}, \quad \wp(z,g_{2},g_{3})=\xi,$$
where $\wp$ is the Weierstrass elliptic function, by the following change of variables \cite{Wang_Guo}
$$z=-2u+\frac{1}{2}.$$
The integral then transforms into the form
$$I=-\int_{\infty}^{u_{0}} \frac{du}{\sqrt{4u^{3}-\frac{3}{4}u- \frac{1}{4}(-C-\frac{1}{2})}}.$$
Hence, by the fact that the Weierstrass function is even, the solution is of the form
\begin{equation}
 \theta(x)=4x^{-2}\left(\frac{1}{2}- 2\wp \left( \frac{2}{\sqrt{3}} \ln(Bx),\frac{3}{4},-\frac{1}{4} \left( C + \frac{1}{2} \right) \right) \right),
\end{equation}
where $B$ is a different integration constant denoted for simplicity by the same letter.

For $C=1$, there is also $z=1$ value for which $w_{62}(1)=0$. This case corresponds to the solution (\ref{b_solution}).

\subsection{$C \in (-1;0)$}
In this case, for $z < a$ the solution is the same as in the proceeding case. For $ z \in [b,c]$ we can use a simple formula to relate the integrals:

$$\int_{\frac{1}{4}-\frac{b}{2}}^{u_{0}<\frac{1}{4}-\frac{c}{2}} f(x)dx= \int_{\infty}^{u_{0}} f(x)dx - \int_{\infty}^{\frac{1}{4}-\frac{b}{2}} f(x)dx.$$

The last elliptic integral is in general not real ($w_{62}(u)<0$), therefore this case is not interesting in physical applications.

\subsection{$C=0$}
In this case it can be shown that the solution is the  scaled solution (\ref{Talenti-Aubin_solution}) in the following way.
The equation can be integrated
$$\int \frac{dz}{\sqrt{-2z^{3} + 3 z^{2}}} = \pm \int \frac{2dy}{\sqrt{3}},$$
which gives after using $y=-\ln(x)$
$$z=\frac{3}{2}( 1 - tanh^{2}(\pm \ln(Bx))=\frac{6(Bx)^{2}}{(1+(Bx)^{2})^{2}}.$$
Setting $2\sqrt{3}B=\frac{1}{\lambda}$ and restoring $\theta(x)=4\frac{z}{x^{2}}$ we obtain the scaled solution (\ref{Talenti-Aubin_solution})
$$\theta(x)=\frac{1}{\lambda^{2}}\frac{1}{(1+\frac{1}{24}( x / \lambda)^{2})^{2}}.$$

\section{Conclusions}
The analysis presented in this paper is an extension of the results presented in \cite{Mach_Lane_Emden}. We see that similar features can be found. The most characteristic are the cases when $C=-1$ when the solution (\ref{b_solution}) appears, and the cases $C=0$ when the solution (\ref{Talenti-Aubin_solution}) is recovered. In other cases elliptic integrals are involved. This shows that the critical $d$, $p$ cases are quite similar to each other.

\begin{figure}[htp]
\centering
 \includegraphics[scale=0.7]{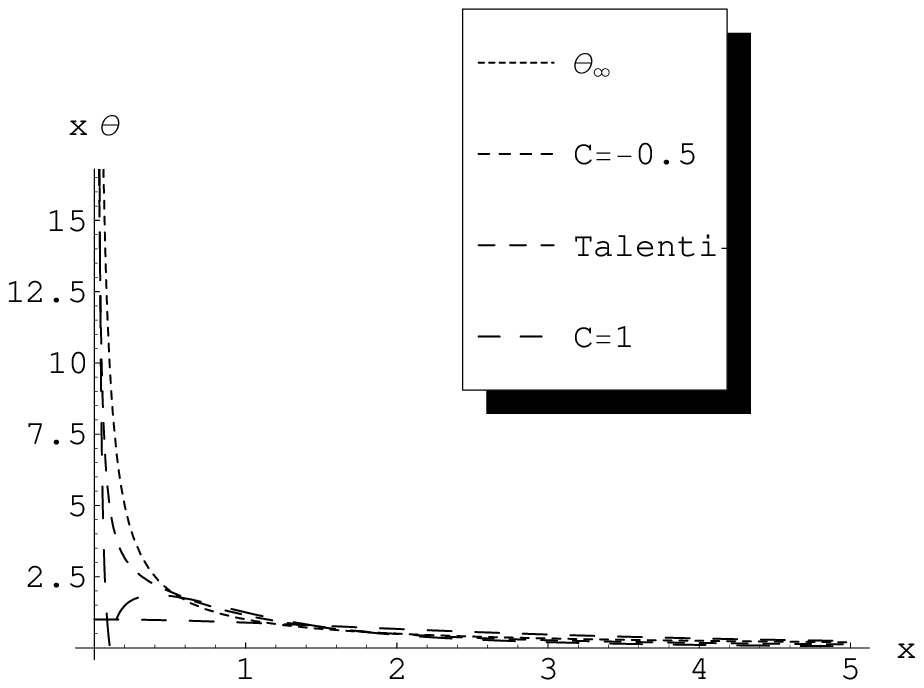}
 \includegraphics[scale=0.7]{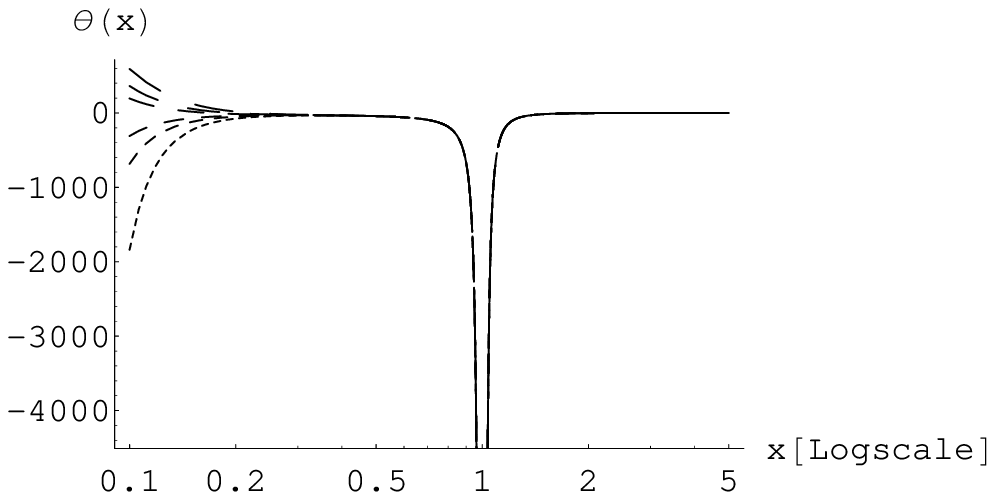}
 \caption{Left panel presents all solutions of $d=4$, $p=3$ Lane-Emden equation. On the right panel there are all solutions of $d=6$, $p=2$ Lane-Emden equation. Solid line - (\ref{b_solution}) and Talenti-Aubin, dashed lines $C=-2$, $C=-1$, $C=-0.5$, $C=1$, $C=2$. Logarithmic scale on $0X$ axis is used. }
 \label{rysunek_43_62_all_solutions}
\end{figure}

Plots for $d=4$, $p=3$ and $d=6$, $p=2$ cases are presented in Figure \ref{rysunek_43_62_all_solutions}.
%
%

One can also observe that when the space dimension $d$ increases then $p$ related by (\ref{critical_p}) decreases, and the polynomial order of the right-hand side of (\ref{critical_equation_integrated}) decreases therefore some solutions that are present in $d=3$, $p=5$, namely the Srivastava solutions \cite{Mach_Lane_Emden}, are missing.

\section{Acknowledgements}

RK is supported by the Warsaw Center of Mathematics and
Computer Science from the founds of the Polish Leading National Research
Centre (KNOW). RK is grateful to Patryk Mach for drawing his attention to paper \cite{Mach_Lane_Emden}.
GF is supported by NCN grant 2011/03/B/ST1/00330.

\par

\end{document}